\documentclass[twoside,english]{jfm}
\usepackage[T1]{fontenc}
\usepackage[latin9]{inputenc}
\usepackage[active]{srcltx}
\usepackage{verbatim}
\usepackage{float}
\usepackage{textcomp}
\usepackage{amsmath}
\usepackage{amssymb}
\usepackage{wasysym}
\usepackage{graphicx}
\usepackage{esint}

\makeatletter
\numberwithin{equation}{section}
\numberwithin{figure}{section}

\usepackage{graphicx}
\usepackage{natbib}

\ifCUPmtlplainloaded \else
  \checkfont{eurm10}
  \iffontfound
    \IfFileExists{upmath.sty}
      {\typeout{^^JFound AMS Euler Roman fonts on the system,
                   using the 'upmath' package.^^J}%
       \usepackage{upmath}}
      {\typeout{^^JFound AMS Euler Roman fonts on the system, but you
                   dont seem to have the}%
       \typeout{'upmath' package installed. JFM.cls can take advantage
                 of these fonts,^^Jif you use 'upmath' package.^^J}%
      }
  \else
  \fi
\fi

\ifCUPmtlplainloaded \else
  \checkfont{msam10}
  \iffontfound
    \IfFileExists{amssymb.sty}
      {\typeout{^^JFound AMS Symbol fonts on the system, using the
                'amssymb' package.^^J}%
       \usepackage{amssymb}%
         
         \let\geq=\geqslant
      }{}
  \fi
\fi

\ifCUPmtlplainloaded \else
  \IfFileExists{amsbsy.sty}
    {\typeout{^^JFound the 'amsbsy' package on the system, using it.^^J}%
     \usepackage{amsbsy}}
    {\providecommand\boldsymbol[1]{\mbox{\boldmath $##1$}}}
\fi

\newsavebox{\astrutbox}
\sbox{\astrutbox}{\rule[-5pt]{0pt}{20pt}}

\title[Hopf fibrations for pipe flows]{Hopf fibrations for turbulent pipe flows}

\author[F. Fedele, O. Abessi and P. J. Roberts]%
{F. Fedele$^{1,2}$%
  \thanks{Email address for correspondence: fedele@gatech.edu},\ns
O. Abessi$^1$ and P. J. Roberts$^1$}

\affiliation{$^1$School of Civil and Environmental Engineering, Georgia Institute of Technology,
Atlanta, GA 30322, USA\\[\affilskip]
$^2$School of Electrical and Computer Engineering, Georgia Institute of Technology, Atlanta, GA 30322, USA}

\pubyear{2010}
\volume{650}
\pagerange{119--126}
\date{?; revised ?; accepted ?. - To be entered by editorial office}

\makeatother

\usepackage{babel}
\begin{document}
\maketitle\global\long\def\S{\mathcal{S}}
\global\long\def\eps{\varepsilon}
\global\long\def\H{\mathcal{H}}
\global\long\def\L{\mathcal{L}}
\global\long\def\M{\mathcal{M}}
\global\long\def\K{\mathbf{K}}
\global\long\def\Hilb{\mathbf{H}}
\global\long\def\R{\mathbb{R}}
\global\long\def\ud{\mathrm{d}}

\begin{abstract}
We propose a generalization of Hopf fibrations to quotient the streamwise
translation symmetry of turbulent pipe flows viewed as dynamical systems.
In particular, we exploit the geometric structure of the associated
high dimensional state space, which is that of a principal fiber bundle.
The relation between the comoving frame velocity $U_{d}$ associated
with the dynamical phase of an orbit in the bundle and the Taylor's
hypothesis is investigated. As an application, Laser-Induced-Fluorescence
techniques are exploited to capture planar fluorescent dye concentration
fields tracing a turbulent pipe flow at the bulk Reynolds number $\mathfrak{\mathsf{Re}}=3200$.
The symmetry reduction analysis of the experimental data reveals that
the speed $u$ of dye concentration bursts is associated with the
dynamical and geometric phases of the corresponding orbits in the
fiber bundle. In particular, in the symmetry-reduced frame we unveil
a pattern-changing dynamics of the passive scalar structures, which
explains the observed speed $u\approx U_{d}+U_{g}$ of intense bursting
events in terms of the geometric phase velocity $U_{g}\approx0.43U_{d}$
associated with the orbits in the bundle.\end{abstract}

\section{Introduction}

In the last decade, Navier-Stokes turbulence in channel flows has
been studied as a chaotic dynamics in the state space of a high-dimensional
system at moderate Reynolds numbers (see, for example, \cite{GibsonetalJFMCouette2008,WillisCvitanovic2013}).
Here, turbulence is viewed as an effective random walk in state space
through a repertoire of invariant solutions of the governing equations
(\cite{CvitanovicJFM2013_clockwork} and references therein). In state
space, turbulent trajectories visit the neighbourhoods of equilibria,
travelling waves or periodic orbits, switching from one saddle to
the other through their stable and unstable manifolds (\cite{CvitanovicPOT_1991},
see also \cite{ChaosBook}). Recent studies on the geometry of the
state space of Kolmogorov flows (\cite{Chandler_Kerswell2013_Kolm})
and barotropic atmospheric models (\cite{Gritsun2011,Gritsun2013})
give evidence that unstable periodic orbits provide the skeleton underpinning
the chaotic dynamics of fluid turbulence. 

In pipe flows, the intrinsic continuous streamwise translation symmetry
and azimuthal discrete symmetry make difficult identifying invariant
flow structures, such as traveling waves or relative equilibria (\cite{FaisstEckhardt,WedinKerswell2004})
and relative periodic orbits (\cite{VISWANATH2007}), embedded in
turbulence. These travel downstream with their own mean velocity and
there is no unique comoving frame that can simultaneously reduce all
relative periodic orbits to periodic orbits and all travelling waves
to equilibria. Recently, this issue has been addressed by \cite{WillisCvitanovic2013}
using the method of slices (\cite{Siminos2011,Froehlich}, see also
\cite{RowleyMarsden2000,RowleyMarsden2003}) to quotient group symmetries
and reveal the geometry of the state space of pipe flows at moderate
Reynolds numbers. Further, \cite{Budanur2014} exploit the 'first
Fourier mode slice' to reduce the $SO(2)$-symmetry in spatially extended
systems. In particular, they separate the dynamics of the Kuramoto-Shivasinsky
equation into a `shape-changing' dynamics within a quotient or symmetry-reduced
space (base manifold) and a one-dimensional (1-D) transverse space
(fiber) associated with the group symmetry. This is the geometric
structure of a principal fiber bundle (e.g. \cite{Husemoller}), the
main topic of our work. In particular, we propose a generalization
of Hopf fibrations (\cite{Hopf1931}, see also \cite{Steenrod}) for
dynamical systems with translation symmetries, and apply it to symmetry-reduce
the evolution of passive scalars of turbulent pipe flows. 

The paper is organized as follows. We first discuss the limitations
of the method of comoving frames, also referred to as the method of
connections (e.g. \cite{RowleyMarsden2000}). Then, we provide an
overview of principal fiber bundles and present the symmetry reduction
scheme via Hopf fibrations. As an application, two-dimensional (2-D)
Laser-Induced-Fluorescence (LIF) techniques are exploited to capture
planar fluorescent dye concentration fields tracing turbulent pipe
flow patterns at Reynolds number $\mathsf{Re}=2U_{b}R/\nu=3200$,
where $U_{b}$ is the bulk velocity, $R$ is the radius and $\nu$
is the kinematic viscosity of water. Symmetry reduction of the acquired
experimental data is then presented and discussed.

\section{Comoving frame velocities and Taylor's hypothesis}

Consider a 2-D passive scalar field $C(x,y,t)$ advected by a velocity
field $\mathbf{u}(x,y,t)=(U,V)$ and that evolves according to

\begin{equation}
\partial_{t}C+\mathbf{u}\cdot\nabla C=d\nabla^{2}C+f-s,\label{1}
\end{equation}
where $d$ is the diffusion coefficient, $f$ and $s$ are sources
and sinks, and ($x,y$) are the horizontal streamwise and vertical
cross-stream directions, respectively. The generalization to three-dimensional
geometries is straightforward, and it will not be discussed here.
Assume that (\ref{1}) admits streamwise translation symmetry, that
is if $C(x,y,t)$ is a solution so is $C(x-x_{d},y,t)$ for any drift
$x_{d}$. This can be time-varying and related to a comoving frame
velocity $U_{d}=\frac{dx_{d}}{dt}$, for which the material derivative

\begin{equation}
\frac{DC}{Dt}=\partial_{t}C+U_{d}\partial_{x}C
\end{equation}
is, in average, the smallest possible, namely

\begin{equation}
\left\langle \left(\partial_{t}C+U_{d}\partial_{x}C\right)^{2}\right\rangle _{x,y}
\end{equation}
is minimal if 
\begin{equation}
U_{d}(t)=-\frac{\left\langle \partial_{t}C\partial_{x}C\right\rangle _{x,y}}{\left\langle \left(\partial_{x}C\right)^{2}\right\rangle _{x,y}},\label{Uc}
\end{equation}
where the brackets $\left\langle \cdotp\right\rangle {}_{x,y}$ denote
space average in $x$ and $y$. In the comoving frame $\left(x-U_{d}t,t\right)$,
the passive scalar appears to flow 'calmly', but still slowly drifting
(see, for example, \cite{KreilosJFM2014} for a study of Couette flows).
Only when $\frac{DC}{Dt}=0$, namely diffusion, sources and sinks
balance, the flow is steady in the comoving frame (\cite{KrogstadPoF}),
as for travelling waves (\cite{FaisstEckhardt,WedinKerswell2004}).
From (\ref{1}), (\ref{Uc}) can be written as 
\begin{equation}
U_{d}(t)=\frac{\left\langle U\left(\partial_{x}C\right)^{2}+V\partial_{x}C\partial_{y}C-(f-s)\partial_{x}C\right\rangle _{x,y}}{\left\langle \left(\partial_{x}C\right)^{2}\right\rangle _{x,y}},
\end{equation}
which reveals that the comoving frame velocity is a weighted average
of the local flow velocities, sources and sinks. Although diffusion
processes are invariant under translation, they indirectly affect
$U_{d}$ through the concentration gradients. From (\ref{Uc}), averaging
along the $x$ direction only yields the vertical comoving frame velocity
profile 
\begin{equation}
U_{d}(y,t)=-\frac{\left\langle \partial_{t}C\partial_{x}C\right\rangle _{x}}{\left\langle \left(\partial_{x}C\right)^{2}\right\rangle _{x}}.\label{Ucpr}
\end{equation}
 The comoving frame speed $\widehat{U}_{d}$ of a Fourier mode $\widehat{C}(k,y,t)e^{ikx}$
then follows as

\begin{equation}
\widehat{U}_{d}(k,y,t)=-\mathrm{Im}\frac{\partial_{t}\widehat{C}(k,y,t)\overline{\widehat{C}}(k,y,t)}{k\left|\widehat{C}(k,y,t)\right|^{2}},\label{Uck}
\end{equation}
where $\overline{\widehat{C}}$ is the complex conjugate of $\widehat{C}$
and $\mathrm{Im}(a)$ denotes the imaginary part of $a$. Note that
$\widehat{U}_{d}$ is the same as the convective turbulent velocity
formulated by \cite{Alamo_JImenez2009} in the context of Taylor\textquoteright s
(1938) \nocite{Taylor1938} abstraction of turbulent flows as fields
of frozen eddies advected by the mean flow. If the turbulent fluctuation
$u$ is small compared to the mean flow speed $U_{m}$, then the temporal
response at frequency $\omega$ at a fixed point in space can be viewed
as the result of an unchanging spatial pattern of wavelength $2\pi/k$
convecting uniformly past the point at velocity $U_{m}$, viz. $U_{m}=\omega/k$.
This is the Taylor's hypothesis that relates the spatial and temporal
characteristics of turbulence. However, eddies can deform and decay
as they are advected by the mean flow and their speed may differ significantly
from $U_{d}$. 

In this regard, \cite{Alamo_JImenez2009} concluded that the comoving
frame velocity $U_{d}$ of the largest-scale motion is close to the
local mean speed $U_{m}$, whereas $U_{d}$ drops significantly as
the scales are reduced (\cite{KrogstadPoF}). Hence, there is no unique
convection velocity, which insteads depends upon the state of evolution
of the flow. For example, it is well known that the turbulent motion
in channel flows is organized in connected regions of the near wall
flow that decelerate and then erupt away from the wall as \textquotedbl{}ejections\textquotedbl{}.
These decelerated motions are followed by larger scale connected motions
toward the wall from above as \textquotedbl{}sweeps\textquotedbl{}.
\cite{KrogstadPoF} found that the convection velocity for ejections
is distinctly lower than that for sweeps. 

To gain more insights into the physical meaning of comoving frame
velocities, we have performed experiments to trace turbulent pipe
flow patterns exploiting non-intrusive LIF techniques (\cite{TianRoberts2003})
as discussed below.

\subsection{LIF measurements }

\begin{figure}[H]
\centering \includegraphics[width=0.99\textwidth]{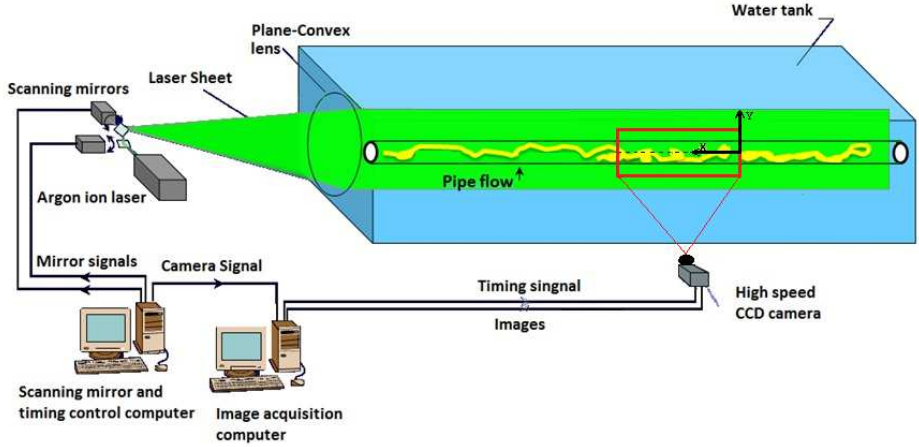}
\protect\caption{Schematic of the LIF system of the Georgia Tech Environmental Fluid
Mechanics Laboratory (\cite{TianRoberts2003}). }

\label{FIGURE1} 
\end{figure}

\begin{figure}[H]
\centering \includegraphics[width=1\textwidth]{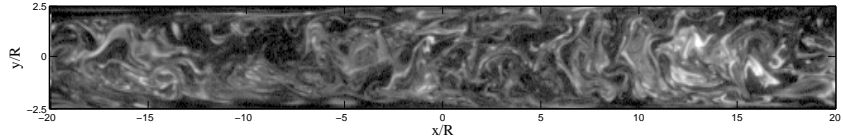} \protect\caption{LIF experiments: snapshot of the planar fluorescent dye concentration
field $C(x,y,t)$ (log scale) tracing turbulent pipe flow patterns
at Reynolds number $\mathsf{Re}=3200$ (bulk velocity $U_{b}=6.42\mbox{ cm/s}$,
flow from right to left). }

\label{FIGURE1a} 
\end{figure}

The experiments were performed in the Environmental Fluid Mechanics
Laboratory at the Georgia Institute of Technology. The LIF configuration
is illustrated in Fig. (\ref{FIGURE1}) and a detailed description
of the system is given in \cite{TianRoberts2003}. The tank has glass
walls $6.10$ m long \texttimes{} $0.91$ m wide \texttimes{} $0.61$
m deep. The front wall consists of two three-meter long glass panels
to enable long unobstructed view. The $5.5$ meter long pipe located
on the tank floor and tank was filled with water that was filtered
and dechlorinated. The pipe was transparent Perspex tube with radius
$R=2.5\mbox{ cm}$. The tank was filed up to well above the pipe to
avoid reflections of the laser sheet on the pipe wall. The water first
pomp into a damping chamber to get calm. Then, after passing a rigid
polyester air filter, it flowed into the pipe. A volume of fluorescent
dye solution continuously injected into the flow through a small hole
in the pipe wall upstream of the capture zone of length of $20R$.
The solution, a mixture of water and fluorescent dye, is supplied
from a reservoir by a rotary pump at a flowrate measured by a precision
rotameter. The flow was begun and, after waiting few minutes for the
flow to establish, laser scanning started to record the experiment.
To acquire high resolution data, we captured planar fluorescent dye
concentration fields $C(x,y,t)$ tracing turbulent pipe flow patterns
at Reynolds number $\mathsf{Re}=2U_{b}R/\nu=3200$ (bulk velocity
$U_{b}=6.42\mbox{ cm/s}$). As shown in Fig. (\ref{FIGURE1}), a laser
sheet was located at the center of the pipe to focus on flow properties
in the central plane ($y=0$ is the pipe centerline). Images of the
capture zone ($2R\times20R=5\times50\:\mbox{c\ensuremath{m^{2}}}$)
were acquired at $50$ Hz for a duration of $240$ seconds (see Fig.
(\ref{FIGURE1a})). Their vertical and horizontal resolutions are
of $65\times622$ pixels and $0.0794$ cm/pixel.

\begin{figure}[H]
\centering \includegraphics[width=1\textwidth]{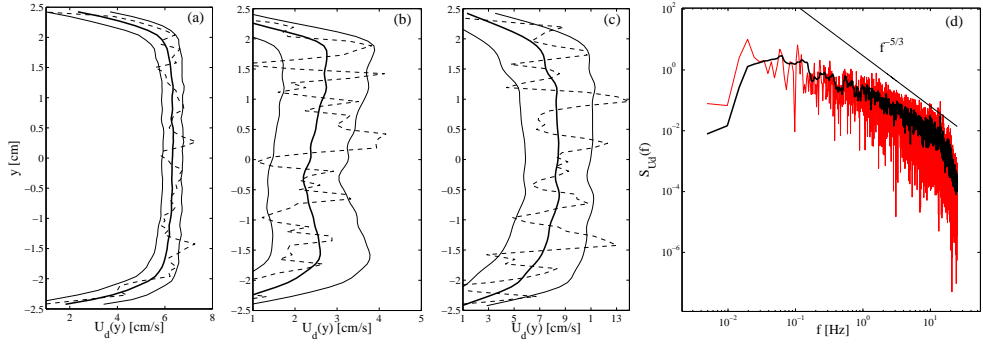}
\protect\caption{Comoving frame velocity profile $U_{d}(y)$ estimated from (a) all
space scales, $U_{d,max}$$=6.32$ cm/s, (b) small scales (wavelengths
$L_{x}<0.2R$, $L_{y}<0.2R$), $U_{d,max}$$=2.76$ cm/s, and (c)
large scales ($L_{x}>2R$, $L_{y}>0.4R$), $U_{d,max}$$=8.52$ cm/s,
and pipe radius $R=2.5$ cm; (d) frequency spectrum of the large-scale
comoving frame velocity $U_{d}$ {[}see Eq. (\ref{Uc}){]}.}

\label{FIGURE2} 
\end{figure}

\begin{figure}[H]
\centering \includegraphics[width=0.8\textwidth]{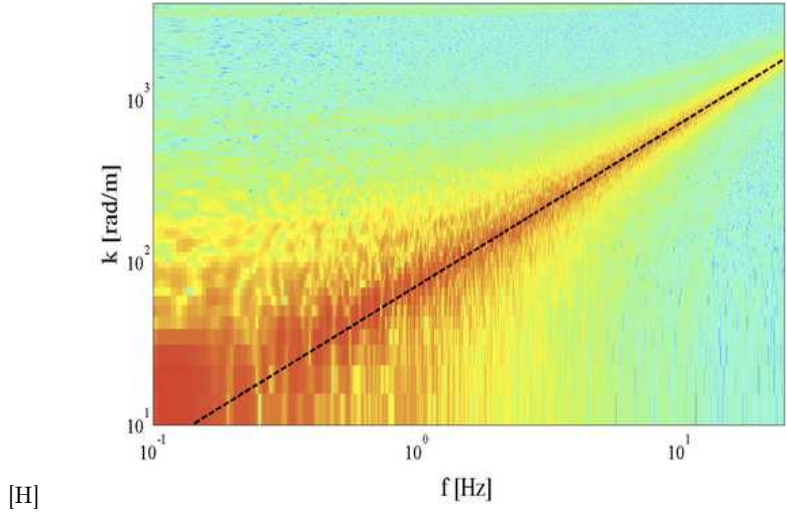}
\protect\caption{Observed frequency-wavenumber spectrum of the fluorescent dye concentration
$C(x,y=0,t)$ at the pipe centerline. Estimated mean velocity $U_{m}=\omega/k\sim8.78$
cm/s (dashed line). $U_{m}/U_{b}=1.37$ and bulk velocity $U_{b}=6.42$
cm/s.}

\label{FIGURE3} 
\end{figure}

\begin{figure}[H]
\centering \includegraphics[width=0.99\textwidth]{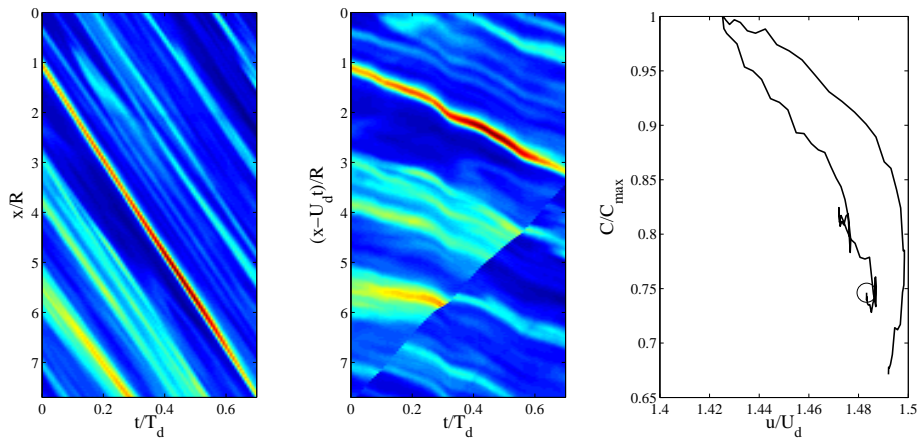}
\protect\caption{LIF experiments: space-time evolution of the dye concentration $C(x,y=0,t)$
at the pipe centerline in the (left) lab frame ($x,t)$ and (center)
comoving frame $\mbox{\ensuremath{(x-U_{d}t,t)}}$; (right) normalized
concentration peak intensity $C/C_{max}$ tracked from the initial
time $t/T_{d}=0$ ($\Circle$) as function of the observed peak speed
$u/U_{d}$, with $C_{max}$ denoting the observed maximum value of
$C$. $U_{d}\approx$6.34 m/s and $T_{d}=U_{d}/R$.}

\label{FIGURE4} 
\end{figure}

\begin{figure}[H]
\centering\includegraphics[width=0.75\columnwidth]{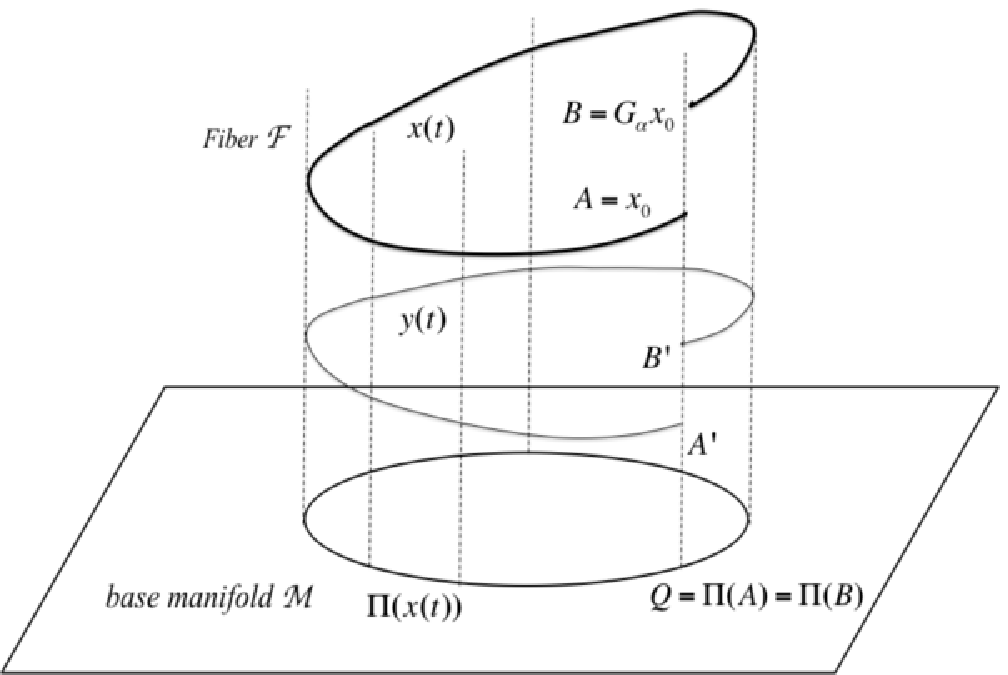}
\protect\caption{Principal fiber bundle: a relative periodic orbit AB reduces to a
periodic orbit in the base manifold $\mathcal{M}$ by properly phase-shifting
the trajectory along the fiber $\mathcal{F}$ (or Lie-group space).
The shift is composed by a dynamical and geometric phases. The shift
induced by the dynamical phase yields the comoving trajectory A'C'B',
which is locally transversal to the fibers (parallel transport through
the fiber bundle), but it is not a closed trajectory. A further shift
by the geometric phase reduces A'C'B' to a periodic orbit on the base
manifold $\mathcal{M}$. }

\label{FIGURE5} 
\end{figure}

\subsection{Data analysis}

The vertical comoving frame velocity profile can be estimated from
the measured fluorescent dye concentration field $C(x,y,t)$ using
Eq. (\ref{Ucpr}). For example, Fig. (\ref{FIGURE2}) shows $U_{d}(y)$
computed accounting for (Panel $a$) all space scales , (Panel \textbf{$\mathit{b}$})
small scales (wavelengths $L_{x}<0.2R$, $L_{y}<0.2R$) and (Panel
$c$) large scales of $C$ ($L_{x}>2R$, $L_{y}>0.4R$). Clearly,
small scales advect at lower speed than larger scales, in agreement
with \cite{KrogstadPoF}. Moreover, the maximum convective velocity
of large scales ($=8.52\mbox{ cm/s}$) is close to the centerline
mean flow speed (=$8.78$ cm/s) estimated from the frequency-wavenumber
spectrum of $C(x,y=0,t)$ {[}see Figure (\ref{FIGURE3}){]}. The frequency
spectrum of $U_{d}(t)$ estimated from Eq. (\ref{Uc}) accounting
for large scales only is also shown in Panel $d$ of Fig. (\ref{FIGURE2}).
It decays approximately as $f^{-5/3}$ as an indication that the Taylor's
hypothesis is approximately valid, possibly due to the non-dispersive
behavior of large scale motions.

In the lab frame $(x,t)$, the space-time evolution of the measured
fluorescent dye concentration $C(x,y=0,t)$ at the pipe centerline
is shown in the left panel of Figure (\ref{FIGURE4}). The associated
evolution in the comoving frame $\left(x-U_{d}t,t\right)$ is also
shown in the center panel. Here, $U_{d}$ is estimated from Eq. (\ref{Uc})
accounting for all space scales of $C$. Note the pattern-changing
dynamics of the passive scalar structures, which still experience
a drift in the comoving frame. Moreover, a generic slowdown or decelerated
motion is observed as the dye concentration bursts, possibly related
to the abovementioned turbulent flow ejections. This is clearly seen
in the the right panel of the same Figure, which reports the normalized
concentration peak intensity $C/C_{max}$ as function of the observed
peak speed $u/U_{d}$, with $C_{max}$ denoting the maximum value
of $C$. Further, $u$ is approximately 40-50\% larger than the comoving
frame velocity $U_{d}$, which is also roughly constant during the
event ($U_{d}=6.32\pm0.22$ cm/s). Note that in oceanic wave groups,
large focusing crests tend to slow down as they evolve within the
group, as a result of the natural wave dispersion of unsteady wave
trains (\cite{Banner_PRL2014,JFMFedele2014,FedeleEPL2014}). Thus,
we argue that the observed slowdown of the passive scalar bursts may
be due to the wave dispersive nature of small-scale turbulent structures.

In the following, drawing from differential geometry we will study
the fiber bundle structure of the state space associated with the
fluorescent dye concentration field evolution, which allows to explain
the observed excess speed $u-U_{d}$ of concentration bursts in terms
of geometric phases of the orbits in the bundle.

\section{Principal fiber bundles}

The geometric structure of the state space $\mathcal{P}\in\mathbb{R}^{N}$
of a dynamical system with a continuous Lie-group symmetry $G_{\alpha}$
and parameter $\alpha\in\mathbb{\mathcal{\mathbb{R}}}$, is that of
a principal fiber bundle: a base manifold $\mathcal{M}$ of dimension
$N-1$ (quotient space) and one-dimensional (1-D) fibers attached
to any point of $\mathcal{M}$ (e.g. \cite{Steenrod,Husemoller}).
The fiber $\mathcal{F}$ is the subspace of the Lie group orbit $G_{\alpha}(z)$.
One can think of the group as being an action, which pushes points
in the bundle around the bundle along the fibers (see Fig. \ref{FIGURE5}).
For example, the Euclidean space $\mathbb{R}^{3}$ can be seen as
a fiber bundle of parallel straight lines. The base manifold is a
plane cutting the whole set of parallel lines. This is a trivial fibration
since the total space $\mathcal{P}$ is both locally and globally
the direct product of the base $\mathcal{M}$ and the fiber $\mathcal{F}$,
that is $\mathbb{R}^{3}=\mbox{\ensuremath{\mathcal{M}\text{\texttimes}}\ensuremath{\mathbb{\mathcal{F}=}}}\mathbb{R}^{2}\times\mathbb{R}$.
A famous non trivial fiber bundle is the Hopf fibration of $S^{3}$
spheres by great circles $S^{1}$ and base space $S^{2}$ (\cite{Hopf1931},
see also \cite{Steenrod}). The Hopf fibration, like any fiber bundle,
is locally a product space, i.e. $S^{3}=S^{2}\times S^{1}$, but not
globally. There are numerous generalizations of it. For example, the
unit sphere $S^{2n+1}$ in the complex space $\mathbb{C}^{n+1}$ fibers
naturally over the complex projective space $\mathbb{C}P^{n}$ with
circles as fibers (see appendix). 

A principal fiber bundle is denoted with the quadruplet $(\mathcal{P},\mathcal{M},G_{\alpha},\Pi)$
with total space $\mathcal{P}$ over the base manifold $\mathcal{M}$,
and a Lie group $G_{\alpha}$. The map $\Pi:\mathcal{P}\overset{}{\rightarrow}\mathcal{M}$
projects an element $z$ of the state space $\mathcal{P}$ and all
the elements of the group orbit $G_{\alpha}(z)$ into the same point
$\Pi(z)$ of the base manifold $\mathcal{M}$, viz. $\Pi(z)=\Pi(G_{\alpha}z)$,
with $\alpha\in\mathbb{R}$. In $\mathcal{P}$, a trajectory or orbit
$z(t)$ can be observed in a special comoving frame, within which
the motion is an horizontal transport through the fiber bundle, that
is the comoving orbit $y=G_{\alpha_{d}}z$ is locally transversal
to the fibers (see Fig. \ref{FIGURE5}). The proper shift $\alpha_{d}$
along the fibers to bring the motion in the comoving frame is called
dynamical phase. This increases with the time spent by the trajectory
to wander around $\mathcal{P}$ and system\textquoteright s answer
to: \textquotedblright{} how long did your trip take? \textquotedblright{}
(\cite{Berry08031984}). For example, the translational shift induced
by the constant speed of traveling waves, or relative equilibria,
is the dynamical phase. They reduce to equilibria within the base
manifold $\mathcal{M}$, whereas relative periodic orbits reduce to
periodic orbits (see Fig. \ref{FIGURE5}). In this case, the shift
along the fibers includes also a geometric phase $\alpha_{g}$, induced
by the projected motion within the $\mathcal{M}$ (\cite{Pancharatnam,Simon,Aharonov_Anandan,Garrison_GeomPhases}).
This phase is independent of time and it depends only upon the curvature
of $\mathcal{M}$, and system\textquoteright s answer to:\textquotedblright{}
where have you been? \textquotedblright{} (\cite{Berry08031984}). 

Geometric phases arise due to anholonomy, that is global change without
local change. The classical example is the parallel transport of a
vector on a sphere. The change in the vector direction is equal to
the solid angle of the closed path spanned by the vector and it can
be described by Hannay's angles (\cite{Hannay}). The rotation of
Foucault\textquoteright s pendulum can be explained by means of such
a anholonomy. \cite{Pancharatnam} discovered this effect for polarized
light, and later on \cite{Berry08031984} rediscovered it for quantum-mechanical
systems. In fluid mechanics, \cite{Shapere1} exploited geometric
phases to explain self-propulsion at low Reynolds numbers. Note that
non-periodic orbits in $\mathcal{M}$ induce also a geometric phase,
that is the motion does not have to be periodic to have geometric
drift (\cite{Anandan_Nature}).

In summary, the total phase associated with any orbital path on the
base manifold is a measure of the motion induced within the fibers
of the bundle by the path. In the following, we will propose an Hopf
bundle for dynamical systems with translation symmetries.

\section{Symmetry reduction via Hopf fibrations}

For the sake of simplicity, consider a 1-D space-periodic passive
scalar field $C(x,t)$, which evolves according to
\begin{equation}
\partial_{t}C=\mathcal{N}(C),\label{ceq}
\end{equation}
where $\mathcal{N\mathrm{(\mbox{\ensuremath{C}})}}$ is a differential
operator of its argument. The extension to higher dimensions is straightforward.
The dynamics admits a continuous translation symmetry, that is if
$C(x,t)$ is a solution, so is $C(x+x_{s},t)$ for any drift $x_{s}$.
It is convenient to express $C$ as the Fourier series

\begin{equation}
\begin{array}[t]{c}
C(x,t)=C_{0}(t)+\frac{1}{2}\sum_{m=1}^{N}z_{m}(t)\exp\left(imk_{0}x\right)+c.c.=\\
\\
C_{0}(t)+\sum_{m=1}^{N}\left|z_{m}\right|\cos\left(imk_{0}x+\theta_{m}\right),
\end{array}\label{cf}
\end{equation}
where $C_{0}(t)$ is the mean, $z_{m}=\left|z_{m}\right|\exp(i\theta_{m})$
are complex Fourier amplitudes with phases $\theta_{m}$ and $k_{0}=2\pi/L_{0}$,
with $L_{0}$ denoting the domain length. The mean $C_{0}$ is invariant
under the group action and it can be decoupled from the vector $z(t)=\{z_{m}\}=\left(z_{1},...z_{N}\right)$.
This satisfies the dynamical system

\begin{equation}
\frac{dz}{dt}=\mathcal{\widehat{N}}(z,C_{0}),\label{ODE}
\end{equation}
coupled with 
\begin{equation}
\frac{dC_{0}}{dt}=\left\langle \mathcal{N}(C_{0},z)\right\rangle _{x},\label{ceq-1}
\end{equation}
the space average of (\ref{ceq}). The vector

\begin{equation}
\mathcal{\widehat{N}}(z,C_{0})=\left\{ \widehat{\mathcal{N}}_{m}(z,C_{0})\right\} =\left(\widehat{\mathcal{N}}_{1}(z,C_{0}),...,\widehat{\mathcal{N}}_{N}(z,C_{0})\right),\label{ODE-1-1}
\end{equation}
follows from the discrete Fourier transform of $\mathcal{N}$ in Eq.
(\ref{ceq}). 

The orbit $z(t)$ wanders in the state space $\mathcal{P}\in\mathbb{C^{\mathrm{\mathit{N}}}}$,
and the symmetry group $G_{x_{s}}$ of $z$ is the subspace 

\begin{equation}
G_{x_{s}}(z)=\left\{ w\in\mathbb{C^{\mathit{N}}}:w=\{z_{m}\exp(imk_{0}x_{s})\},\:\forall x_{s}\in\mathbb{R}\right\} .
\end{equation}
The state space $\mathcal{P}$ is geometrically a principal fiber
bundle: a base manifold $\mathcal{M}$ of dimension $2N-1$ (quotient
space) and one dimensional (1-D) fibers $G_{x_{s}}(Z)$ attached to
any point $Z\in\mathcal{M}$ (e.g. \cite{Steenrod,Husemoller}). The
bundle is described by the quadruplet $(\mathcal{P},\mathcal{M},G_{x_{s}},\Pi_{j})$
and the map $\Pi_{j}$ is given as follows. For a non-vanishing Fourier
mode $z_{j}$, the desymmetrized orbit $z_{D}(t)$ within the base
manifold $\mathcal{M}\in\mathbb{C^{\mathit{N}}}$ is given by the
map $\Pi_{j}:\mathit{\mbox{\mbox{\ensuremath{\mathcal{\mathrm{\mathcal{P}}}}}}}\rightarrow\mathcal{M}$ 

\begin{equation}
z_{D}=\Pi_{j}(z)=Z=(Z_{1},...Z_{m},...Z_{N}),\label{ZD}
\end{equation}
where the complex amplitudes
\begin{equation}
Z_{m}=\left|z_{m}(t)\right|\exp(i\phi_{m}(t)),
\end{equation}
and phases 

\begin{equation}
\phi_{m}=\theta_{m}-\frac{m\theta_{j}}{j}.
\end{equation}
Note that $Z_{j}$ is real and $\mathcal{M}$ is a $2N-1$ dimensional
manifold of $\mathbb{C}^{N}$. For $j>1$, the phase $\theta_{j}$
needs to be unwrapped in order to avoid introducing spurious discrete
symmetries. For $j=1$, the fibration reduces to the 'first Fourier
mode slice' proposed in \cite{Budanur2014}. The map $\Pi_{j}$ projects
an element $z=\{z_{m}\}$ of the state space $\mathcal{P}$ and all
the elements of its group orbit $G_{x_{d}}(z)$ into the same point
$z_{D}=\Pi_{j}(z)$ in $\mathcal{M}$, namely 
\begin{equation}
\Pi_{j}(z)=\Pi_{j}(G_{x_{s}}(z))\mathrm{.}
\end{equation}
The scalar field $C_{D}$ in the symmetry-reduced frame follows from
(\ref{cf}) as

\begin{equation}
C_{D}(x,t)=C_{0}(t)+\sum_{m=1}^{N}\left|Z_{m}\right|\cos\left(mk_{0}x+\phi_{m}\right).\label{CD}
\end{equation}
Note that $\Pi_{j}$ can be written in the equivalent form 
\begin{equation}
\begin{array}[t]{c}
z_{D}=\Pi_{j}(z)=\ensuremath{\left\{ z_{m}\left(\frac{\left|z_{j}\right|}{z_{j}}\right)^{m/j}\right\} }_{m=1,...N}=\\
\\
\left(z_{1}\left(\frac{\left|z_{j}\right|}{z_{j}}\right)^{1/j},...z_{j-1}\left(\frac{\left|z_{j}\right|}{z_{j}}\right)^{(j-1)/j}\mbox{\ensuremath{,}\ensuremath{\left|z_{j}\right|,z_{j+1}}\text{\ensuremath{\left(\frac{\left|z_{j}\right|}{z_{j}}\right)^{(j+1)/j}}\ensuremath{,...}}\ensuremath{z_{N}\left(\frac{\left|z_{j}\right|}{z_{j}}\right)^{N/j}}}\right),
\end{array}\label{map-1}
\end{equation}
which reveals the geometric structure of a Hopf bundle (e.g. \cite{Steenrod}).
In the following, it is shown that the associated base manifold $\mathcal{M}$,
hereafter referred to as $\mathbb{C}T^{n}$, is a generalization of
$\mathbb{C}P^{n}$ projective spaces.

\subsection{$\mathbb{C}T^{n}$ projective spaces }

The complex projective space $\mathbb{C}P^{n}$ is the quotient space
of the unit $S^{2n+1}$ hypersphere 
\begin{equation}
S^{2n+1}=\left\{ z=(z_{1},...z_{n+1})\in\mathbb{C}^{n+1}:\sum_{j=1}^{n+1}\left|z_{j}\right|^{2}=1\right\} 
\end{equation}
with circles $S^{1}$ as fibers under the action of the $U(1)$ group
(see also appendix). The projection $\pi_{H}:S^{2n+1}\rightarrow\mathbb{C}P^{n}$
is called the Hopf map. Since $\mathbb{C}P^{1}=S^{2}$ is the Riemann
sphere, we obtain the classical Hopf bundle $S^{3}\rightarrow S^{2}$
with fiber $S^{1}$ (\cite{Hopf1931}, see also \cite{Steenrod}).
Hereafter, we generalize $\mathbb{C}P^{n}$ spaces to complex manifolds
defined as quotient spaces 
\begin{equation}
\mathbb{C}T^{n}=E^{2n+1}/T(1)
\end{equation}
of the $E^{2n+1}$ hypersurface in $\mathbb{C}^{n+1}$ 
\begin{equation}
E^{2n+1}=\left\{ z=(z_{1},...z_{n+1})\in\mathbb{C}^{n+1}:\sum_{j=1}^{n+1}\left|z_{j}\right|^{\frac{2(n+1)}{j}}=1\right\} 
\end{equation}
under the action of the translation group $T(1)$ 
\begin{equation}
T_{\lambda}(z)=\left\{ Z=(Z_{1},...Z_{n+1})\in\mathbb{C}^{n+1}:Z_{k}=z_{k}\lambda^{k},\quad\left|\lambda\right|=1\right\} .
\end{equation}
An element $z$ of $\mathbb{C}T^{n}$ is identified with the equivalence
class
\[
[z]=[z_{1},...z_{n+1}]\sim(\lambda z_{1},...\lambda^{n+1}z_{n+1}),\qquad\lambda\neq0.
\]
Two points of $\mathbb{C}^{n+1}\left\backslash \left\{ 0\right\} \right.$
are equivalent if they lie on the same subspace or group orbit $T_{\lambda}(z)$.
For $n=1$, the group orbit is the space of complex parabolas $z_{2}=\lambda z_{1}^{2}$
($z_{1}\neq0$) and, for $n>1$, that of complex hypercurves $z_{n}=\lambda z_{j}^{n/j}$
($z_{j}\neq0$). All points of $T_{\lambda}(z)$ reduce to the same
point of $\mathbb{C}T^{n}$ and the projection map $\pi:\mathbb{C}^{n+1}\rightarrow\mathbb{C}T^{n}$
can be defined as follows. 

The $\mathbb{C}T^{n}$ manifold can be covered by an atlas of charts
or slices $U_{j}=\left\{ [z]\left|z_{j}\neq0\right.\right\} \subset\mathbb{C}T^{n}$,
where the complex hyperplane $B_{j}=\left\{ z_{j}=0\right\} $ is
the ``border'' of $U_{j}$ (\cite{Siminos2011}). Then, the projection
onto a chart or slice $U_{j}$ is given by the map $\pi_{j}:\mathbb{C}^{n+1}\rightarrow U_{j}$ 

\begin{equation}
\pi_{j}([z])=\pi_{j}([z_{1},...z_{n+1}])=\left\{ z_{k}\left(\frac{\left|z_{j}\right|}{z_{j}}\right)^{k/j}\right\} _{k=1,...n+1},
\end{equation}
which is the same map as in Eq. (\ref{map-1}) for $n=N-1$. Thus,
the fiber bundle formulated in the previous section is generalization
of a Hopf fibration for dynamical systems with $T(1)$ symmetries.
The associated state space fibrates over a $\mathbb{C}T^{n}$ manifold
with fibers as the hypercurves $T_{\lambda}(z)$. fibrates over a
$\mathbb{C}T^{n}$ manifold with fibers as the hypercurves $T_{\lambda}(z)$.
Within the charts or slices, relative equilibria reduce to equilibria
and relative periodic orbits reduce to periodic orbits. Clearly, when
$z_{j}$ approaches zero, but likely never completely vanishes, the
trajectory in $\mathbb{C}T^{n}$ wanders nearby the border $B_{j}$
of the slice $U_{j}$ and the associated geometric phase tends to
become singular (see, for example, \cite{Budanur2014}). As described
in \cite{Siminos2011}, a different chart can then be chosen and the
charts' borders can be glued together via ridges into an atlas that
spans the state space region of interest. 

\subsection{Dynamical and geometric phases}

The desymmetrized orbit $z_{D}(t)=Z(t)=\left\{ Z_{m}(t)\right\} $
within the base manifold $\mathcal{M}$ is determined by properly
shifting $z(t)=\left\{ z_{m}(t)\right\} $ along the fibers by $x_{s}(t)$,
that is $Z=G_{-x_{s}}(z)$ and components 
\begin{equation}
z_{m}=Z_{m}\exp(imk_{0}x_{s}),\qquad m=1,...N,\label{zn}
\end{equation}
To find $x_{s}$, we impose the condition of transversality to the
fiber or group orbit 
\begin{equation}
\overline{T_{x_{s}}(Z)}\frac{dZ}{dt}=0,\label{orth}
\end{equation}
namely the tangent $\frac{dZ}{dt}$ at $Z$ must be orthogonal to
the group tangent space 
\begin{equation}
T_{x_{s}}(Z)=(G_{x_{s}}^{-1}\partial_{x_{s}}G)Z=\{imk_{0}Z_{m}\}.
\end{equation}
From (\ref{ODE}) and (\ref{zn}), (\ref{orth}) yields 
\[
\overline{T_{x_{s}}(Z)}\left(\frac{dZ}{dt}+\frac{dx_{s}}{dt}T_{x_{s}}(Z)-\widehat{\mathcal{N}}(Z,C_{0})\right)=0,
\]
from which

\begin{equation}
\frac{dx_{s}}{dt}=\frac{dx_{d}}{dt}+\frac{dx_{g}}{dt},
\end{equation}
where 

\begin{equation}
\frac{dx_{d}}{dt}=U_{d}=\frac{\mathrm{Re}\left[\overline{T_{x_{s}}(Z)}\widehat{\mathcal{N}}(Z,C_{0})\right]}{\left|T_{x_{s}}(Z)\right|^{2}}=-\frac{\mathrm{Im}\sum mk_{0}\overline{Z_{m}}\widehat{\mathcal{N}}_{m}(Z,C_{0})}{\sum m^{2}k_{0}^{2}\left|Z_{m}\right|^{2}}\label{dynphase}
\end{equation}
is the velocity associated with the dynamical phase $x_{d}$, and 

\begin{equation}
\frac{dx_{g}}{dt}=U_{g}=-\frac{\mathrm{Re}\left(\overline{T_{x_{s}}(Z)}\frac{dZ}{dt}\right)}{\left|T_{x_{s}}(Z)\right|^{2}}=\frac{\mathrm{Im}\sum mk_{0}\overline{Z_{m}}\frac{dZ_{m}}{dt}}{\sum m^{2}k_{0}^{2}\left|Z_{m}\right|^{2}}\label{geomphase}
\end{equation}
is that associated with the geometric phase $x_{g}$. Here, the sum
is over $m=1,...N$ and $\mathrm{\mathrm{Re}\mbox{\ensuremath{(\mbox{\ensuremath{a}})}},Im}(a)$
denote the real and imaginary parts of $a$. Note that $\overline{T_{x_{s}}(Z)}\widehat{\mathcal{N}}(Z,C_{0})$
and $\left|T_{x_{s}}(Z)\right|^{2}$ in (\ref{dynphase}) are invariant
under the $T(1)$ group action, and $x_{d}$ can also be determined
replacing $Z$ with the orbit $z$ in $\mathcal{P}$, which is usually
known or observable in applications. The total shift or drift

\begin{equation}
x_{s}=x_{d}+x_{g},\label{drift}
\end{equation}
where 

\begin{equation}
x_{d}(t)=\int_{0}^{t}V_{d}d\tau,\qquad x_{g}(t)=\int_{0}^{t}V_{g}d\tau.\label{xdxg}
\end{equation}
For the Hopf bundle given by the map (\ref{ZD}) or equivalently (\ref{map-1}),
the total drift 
\[
x_{s}=-\frac{\theta_{j}}{k_{0}j},\qquad j\geq1.
\]

The dynamical phase $x_{d}$ increases with the time spent by the
trajectory $z(t)$ to wander around $\mathcal{P}$, whereas the geometric
phase $x_{g}$ depends upon the path $\gamma$ associated with the
motion of $Z(t)$ in the base manifold $\mathcal{M}$. Indeed, from
(\ref{geomphase}) $x_{g}$ can be rewritten as the path integral
\begin{equation}
x_{g}=-\int_{\gamma}\frac{\mathrm{Re}\left(\overline{T_{x_{s}}(Z)}dZ\right)}{\left|T_{x_{s}}(Z)\right|^{2}}=\int_{\gamma}\frac{\mathrm{Im}\sum mk_{0}\overline{Z_{m}}dZ_{m}}{\sum m^{2}k_{0}^{2}\left|Z_{m}\right|^{2}}.\label{xg}
\end{equation}
Clearly, the geometric phase is independent of time and it depends
only upon the reduced motion within $\mathcal{M}$. 

It is straightforward to recognize that $U_{d}$ in (\ref{dynphase})
is the comoving frame velocity introduced in (\ref{Uc}) and specialized
for the dynamical system (\ref{ODE}). In particular, if we shift
the orbit $z(t)$ along the fibers by $x_{d}(t)$, we obtain the comoving
frame orbit 
\[
Z_{d}(t)=G_{-x_{d}}(z)=\left\{ z_{m}(t)\exp(-imk_{0}x_{d}(t)\right\} 
\]
that moves through the fiber bundle locally transversal to the fibers.
This motion is referred to as an ``horizontal transport'' through
the fiber bundle. In physical space, this corresponds to an evolution
in the comoving frame $\left(x-U_{d}t,t\right)$ (see Fig. (\ref{FIGURE4})).
In general, $Z_{d}$ still experiences a shift, the geometric $x_{g}$,
if any. Indeed, the projected path $z_{D}=Z$ within the base manifold
$\mathcal{M}$ is obtained by further shifting $Z_{d}$ along the
fibers by $x_{g}$ (see Fig. \ref{FIGURE5}), namely 
\[
Z(t)=G_{-x_{g}}(Z_{d})=G_{-x_{d}-x_{g}}(z)=\left\{ z_{m}(t)\exp\left[-imk_{0}\left(x_{d}+x_{g}\right)\right]\right\} .
\]
In physical space, this corresponds to a pattern-changing dynamics
in the symmetry-reduced frame $\left(x-(U_{d}+U_{g})t,t\right)$ as
discussed later on. Only relative equilibria or traveling waves have
null geometric phase, since their pattern is not dynamically changing
in the base manifold as they reduce to equilibria. In this case the
comoving and symmetry-reduced frames are the same.

\section{Symmetry reduction of LIF data}

\begin{figure}[H]
\centering \includegraphics[width=0.99\textwidth]{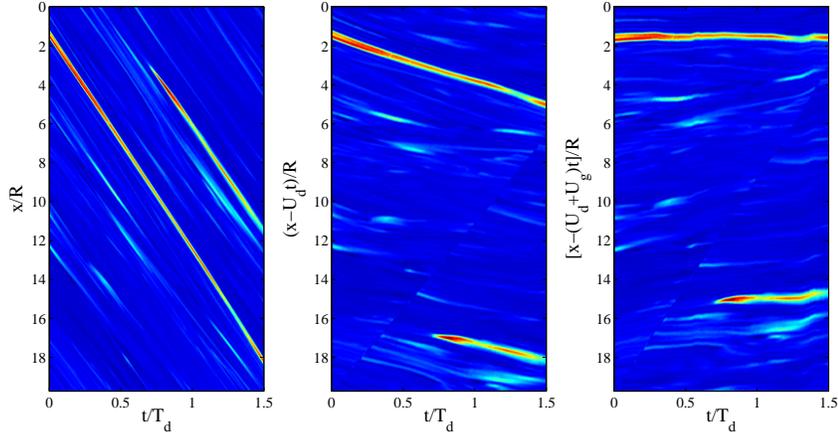}
\protect\caption{Symmetry reduction of LIF data: space-time evolution of a passive
scalar burst event; measured concentration $C(x,y=0,t)$ at the pipe
centerline in the (left) lab frame $(x,t)$, (center) comoving frame
$\left(x-U_{d}t,t\right)$ and (right) symmetry-reduced frame $\left(x-(U_{d}+U_{g})t,t\right)$;
time average $U_{d}\approx6.74$ cm/s, $U_{g}\approx0.4U_{d}$ and
$T_{d}=U_{d}/R$.}

\label{FIGURE6} 
\end{figure}

\begin{figure}[H]
\centering \includegraphics[width=0.99\textwidth]{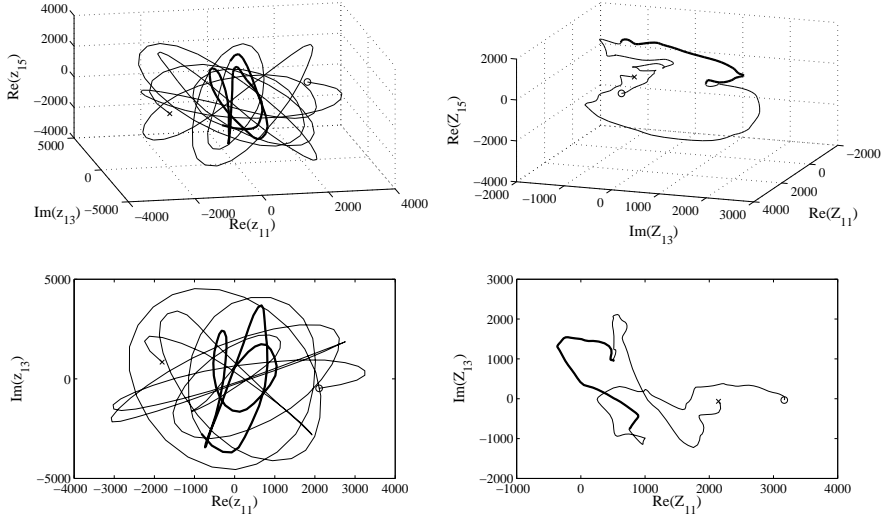}
\protect\caption{Symmetry reduction of LIF data: (left) Orbit trajectories in the subspace
($\mbox{Re}(z_{11}),\mbox{Im}(z_{13}),\mbox{Re}(z_{15})$) of the
state space $\mathcal{P}$ associated with the passive scalar dynamics
in the lab frame of Fig. 7; (right) corresponding symmetry-reduced
orbit within the chart or slice $U_{36}$ of the base manifold $\mathcal{M}$
(subspace ($\mbox{Re}(Z_{11}),\mbox{Im}(Z_{13}),\mbox{Re}(Z_{25})$).
The bold line indicates the excursion of the orbit while the concentration
$C$ lingers above the threshold $0.8C_{max}$ ($\Circle$ =initial
time, $\times$=final time).}

\label{FIGURE7} 
\end{figure}

\begin{figure}[h]
\centering \includegraphics[width=1\textwidth]{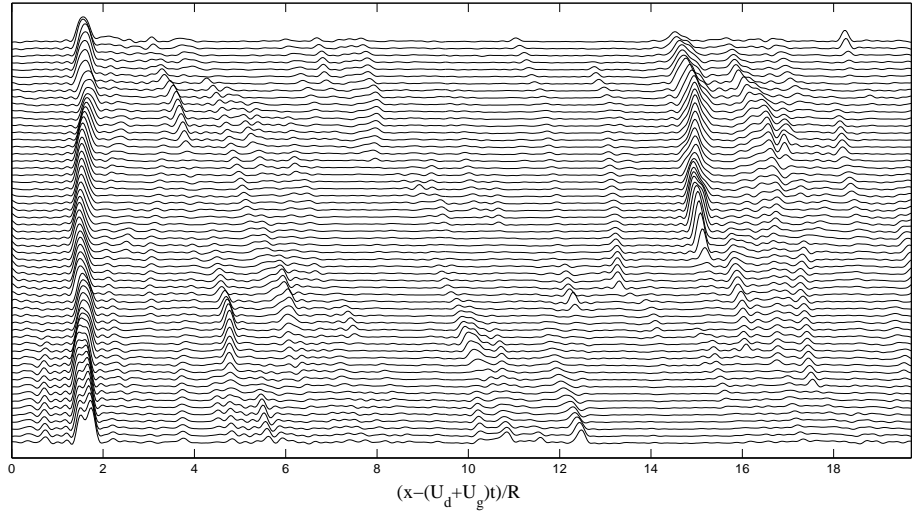}
\protect\caption{Symmetry reduction of LIF data: pattern-changing profiles at increasing
instants of time (from bottom to top) of the measured concentration
$C(x,y=0,t)$ at the pipe centerline in the symmetry-reduced frame
$\left(x-(U_{d}+U_{g})t,t\right)$. Associated 2-D pattern is shown
in the right panel of Fig. (\ref{FIGURE6}).}

\label{FIGURE8} 
\end{figure}

In this section, we apply Hopf fibrations to symmetry-reduce LIF measurements
of turbulent pipe flows at $\mathsf{Re}=3200$ (see section 2.1).
Figure (\ref{FIGURE6}) illustrates the space-time evolution of a
passive scalar burst event. The left panel shows the fluorescent dye
concentration $C(x,y=0,t)$ at the pipe centerline in the lab frame
$(x,t)$. A drift in the streamwise direction $x$ is observed. The
corresponding orbit $z(t)$ in the subspace \{$\mbox{Re}(z_{11}),\mbox{Im}(z_{13}),\mbox{Re}(z_{15})$\}
of the state space $\mathcal{P}$ of dimensions $N=65\times622=40430$
is shown in the left panels of Fig. (\ref{FIGURE7}). Note that the
excursion of the orbit while the concentration $c$ lingers above
the threshold $0.8C_{max}$ is complicated (bold line) since it wanders
around its group orbit $G_{x_{s}}(z)$ as a result of the drift induced
by the translation symmetry. The center panel of Fig. (\ref{FIGURE6})
shows the space-time evolution in the comoving frame $\left(x-U_{d}t,t\right)$.
Note that the dye concentration patterns still experience a significant
drift, the geometric phase $x_{g}$ at velocity $U_{g}$ {[}see Eqs.
(\ref{geomphase}),(\ref{xg}){]}. As a result, the associated orbit
in state space still wanders around the group orbit $G_{x_{s}}(z)$.
The drift disappears in the symmetry-reduced frame $\left(x-(U_{d}+U_{g})t,t\right)$
associated with the dynamics in the chart or slice $U_{36}$ of the
base manifold $\mathcal{M}$, as shown in the right panel of Figure
(\ref{FIGURE6}). Here, one slice is sufficient to symmetry-reduce
the orbit $z$ over the analyzed time span as its component $z_{36}$
never lingers closely near zero. A pattern-changing dynamics of the
passive scalar structures is revealed as clearly seen in Figure (\ref{FIGURE8}).
The corresponding symmetry-reduced orbit $z_{D}(t)=Z(t)$, computed
from Eq. (\ref{ZD}), is shown in the right panels of Fig. (\ref{FIGURE7})
and plotted in the subspace \{$\mbox{Re}(Z_{11}),\mbox{Im}(Z_{13}),\mbox{Re}(Z_{15})$\}
of $\mathcal{M}$. Here, the excursion of the orbit while the concentration
$C$ is high (bold line) indicates an homoclinic type behavior.

\begin{figure}[H]
\centering \includegraphics[width=0.7\textwidth]{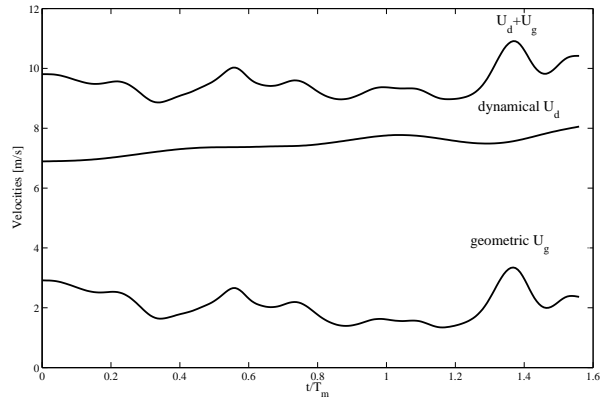}
\protect\caption{Symmetry reduction of LIF data: dynamical and geometric velocities
$U_{d}$ and $U_{g}$ associated with the orbit in state space of
Fig. (\ref{FIGURE7}).}

\label{FIGURE9} 
\end{figure}

\begin{figure}[h]
\centering \includegraphics[width=1\textwidth]{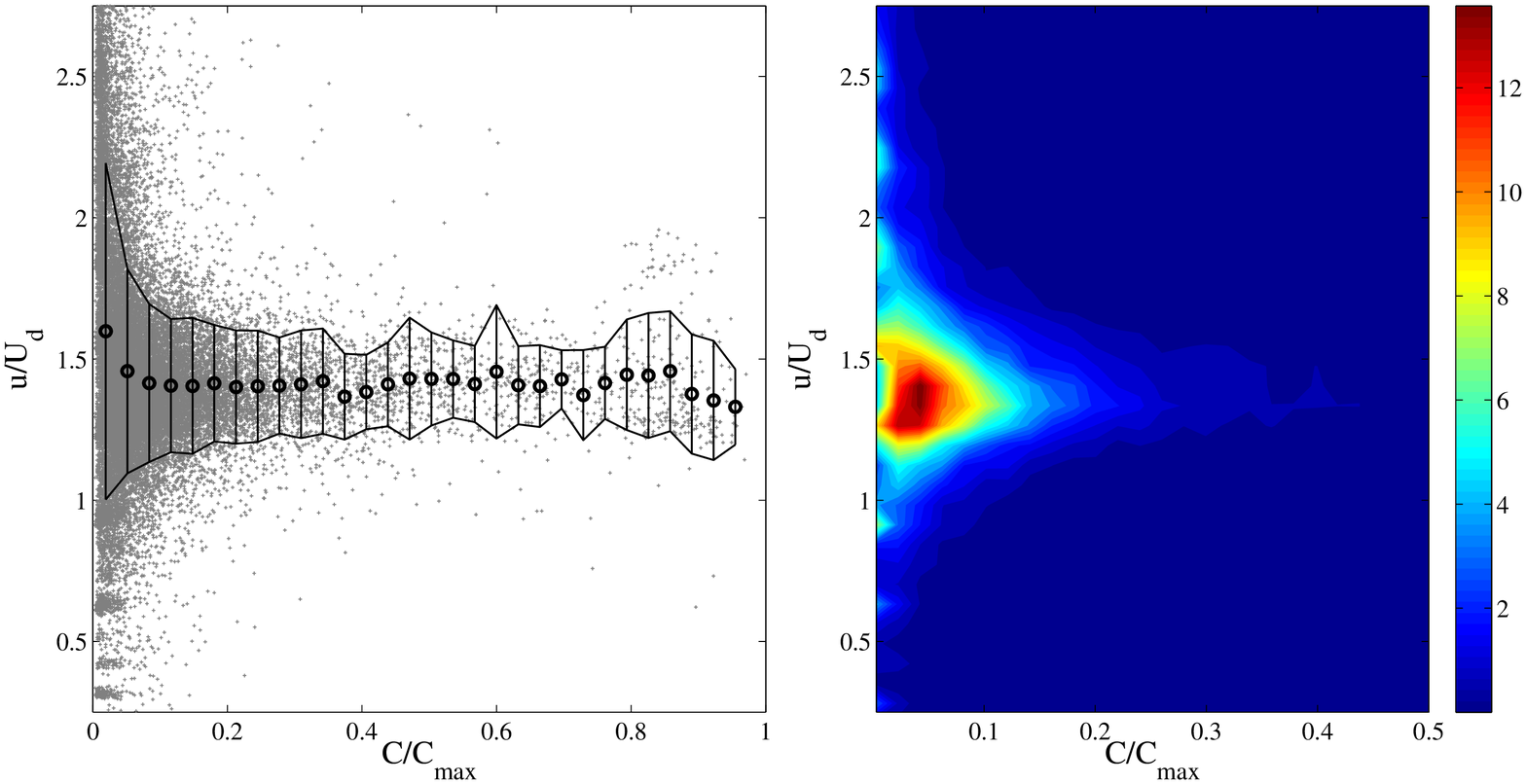}
\protect\caption{LIF experiments: (left) observed normalized dye concentration peak
speed $u/U_{d}$ as function of the amplitude peak $C/C_{max}$, and
(right) associated probability density function. }

\label{FIGURE10} 
\end{figure}

Note that the speed $u\approx U_{d}+U_{g}$ of the dye concentration
burst is approximately 40\% larger than the comoving frame velocity
$U_{d}$, which slighlty changes during the event, whereas the geometric
phase velocity $U_{g}\approx0.4U_{d}$ as seen in Figure (\ref{FIGURE9}).
This appears a generic trend of the flow as clearly seen in Fig. \ref{FIGURE10},
which shows the observed normalized speed $u/U_{d}$ of dye concentration
peaks tracked in space as function of their amplitude $C/C_{max}$,
and the associated probability density function. As the peak amplitude
of concentration bursts increases, their speed $u$ tend to be close
to $1.43U_{d}$. Thus, the space-time evolution in the comoving frame
cannot explain the excess speed $\delta u=u-U_{d}$ (see center panel
of Fig. \ref{FIGURE9}). Indeed, $\delta u$ is roughly the geometric
phase velocity $U_{g}$ induced by the motion in the symmetry-reduced
state space $\mathcal{M}$. In physical space, this results in the
pattern-changing dynamics of the passive scalar structures revealed
in the symmetry-reduced frame (see Figure (\ref{FIGURE8})). Only
when geometric phases are small, viz. $U_{g}\ll U_{d}$, Taylor's
approximation is valid and the flow structures slightly deform as
they are advected at the comoving frame speed $U_{d}$.

\section{Conclusions}

We have presented a theoretical study on the fiber bundle structure
of the state space of dynamical systems with continuous translation
symmetries. A generalization of Hopf fibrations is proposed to quotient
such symmetries in the complex manifold $\mathbb{C}T^{n}$, which
generalizes $\mathbb{C}P^{n}$ projective spaces. As an application,
we have exploited LIF techniques to capture planar fluorescent dye
concentration fields tracing a turbulent pipe flow at Reynolds number
$\mathsf{Re}=3200$. The symmetry reduction of LIF measurements unveils
that the motion of the passive scalar structures is associated with
the dynamical and geometric phases of the corresponding orbits in
the Hopf bundle. In particular, the observed speed $u\approx U_{g}+U_{d}\approx1.43U_{d}$
of dye concentration bursts exceeds the comoving or convective velocity
$U_{d}$. The excess speed $u-U_{d}$ can be explained in terms of
the pattern-changing dynamics in the symmetry-reduced frame and associated
geometric phase velocity $U_{g}\approx0.43U_{d}.$

Hopf fibrations are promising for symmetry reduction of three-dimensional
LIF and PIV measurements as well as simulated flows of pipe turbulence,
in order to unveil the skeleton of its chaotic dynamics in state space.
Further studies along these directions are desirable.

\section{Acknowledgments}

FF acknowledges the Georgia Tech graduate courses ``Classical Mechanics
II'' taught by Prof. Jean Bellissard in Spring 2013 and ``Nonlinear
dynamics: Chaos, and what to do about it?{}`` taught by Prof. Predrag
Cvitanovic in Spring 2012. FF also thanks Prof. Alfred Shapere for
discussions on geometric phases, and Profs. Predrag Cvitanovic, Bruno
Eckhardt and Dr. Evangelos Siminos for discussions on symmetry reduction.

\section{Appendix: $\mathbb{C}P^{n}$ complex projective spaces }

A $\mathbb{C}P^{n}$ space is the set of 1-dimensional complex-linear
subspaces, or lines of the complex space $\mathbb{C}^{n+1}$. In particular,
$\mathbb{C}P^{n}$ is the quotient of $\mathbb{C}^{n+1}\left\backslash \left\{ 0\right\} \right.$
by the equivalent relation
\begin{equation}
z\sim w\Longleftrightarrow\exists\lambda\in\mathbb{C}\left\backslash \left\{ 0\right\} \right.:\: w=\lambda z,\qquad\left|\lambda\right|=1.
\end{equation}
Two points of $\mathbb{C}^{n+1}\left\backslash \left\{ 0\right\} \right.$are
equivalent iff they lie on the same line, viz. they are complex linear
independent. We denote the equivalence class of $z$ by $[z]$, which
is an element of the $U(1)$ group. 

The projection map $\pi:\mathbb{C}^{n+1}\rightarrow\mathbb{C}P^{n}$
can be defined as follows. Consider 

\begin{equation}
z=(z_{1},...z_{n+1})\in\mathbb{C}^{n+1},\qquad z\neq0,
\end{equation}
then $\mathbb{C}P^{n}$ is a complex manifold, which can be covered
by charts $U_{j}=\left\{ [z]\left|z_{j}\neq0\right.\right\} \subset\mathbb{C}P^{n}$,
i.e. the space of all lines not contained in the complex hyperplane
or border $B_{j}=\left\{ z_{j}=0\right\} $. Then, the maps $\pi_{j}:\mathbb{C}^{n+1}\rightarrow U_{j}$
are defined as

\begin{equation}
\begin{array}[t]{c}
\pi_{j}([z])=\pi_{j}([z_{1},...,z_{n+1}])=\left(z_{k}\left(\frac{\left|z_{j}\right|}{z_{j}}\right)\right)_{k=1,...n+1}\\
\\
\left(z_{1}\left(\frac{\left|z_{j}\right|}{z_{j}}\right),...z_{j-1}\left(\frac{\left|z_{j}\right|}{z_{j}}\right),\left|z_{j}\right|,z_{j+1}\left(\frac{\left|z_{j}\right|}{z_{j}}\right),...z_{n+1}\left(\frac{\left|z_{j}\right|}{z_{j}}\right)\right).
\end{array}
\end{equation}
It is straightforward to see that $\mathbb{C}P^{n}$ is the quotient
space of a unit $S^{2n+1}$ sphere under the action of the $U(1)$
group, namely 
\begin{equation}
\mathbb{C}P^{n}=S^{2n+1}/U(1).
\end{equation}
Indeed, every line in $\mathbb{C}^{n+1}$ intersects the unit $S^{2n+1}$
sphere in a circle $S^{1}$, and we obtain a point of $\mathbb{C}P^{n}$
defined by this line by identifying all points on $S^{1}$. %

\bibliographystyle{jfm}
\bibliography{geometricphases_citationstest}

\begin{thebibliography}{35}
\expandafter\ifx\csname natexlab\endcsname\relax\def\natexlab#1{#1}\fi

\bibitem[Aharonov \& Anandan(1987)]{Aharonov_Anandan}
{\sc Aharonov, Y. \& Anandan, J.} 1987 Phase change during a cyclic quantum
  evolution. {\em Phys. Rev. Lett.\/} {\bf 58}, 1593--1596.

\bibitem[Anandan(1992)]{Anandan_Nature}
{\sc Anandan, Jeeva} 1992 The geometric phase. {\em Nature\/} {\bf 360}~(6402),
  307--313.

\bibitem[Banner {\em et~al.\/}(2014)Banner, Barthelemy, Fedele, Allis,
  Benetazzo, Dias \& Peirson]{Banner_PRL2014}
{\sc Banner, M.\, L., Barthelemy, X., Fedele, F., Allis, M., Benetazzo, A.,
  Dias, F. \& Peirson, W.\, L.} 2014 Linking reduced breaking crest speeds to
  unsteady nonlinear water wave group behavior. {\em Phys. Rev. Lett.\/} {\bf
  112}, 114502.

\bibitem[Berry(1984)]{Berry08031984}
{\sc Berry, M.~V.} 1984 Quantal phase factors accompanying adiabatic changes.
  {\em Proceedings of the Royal Society of London. A. Mathematical and Physical
  Sciences\/} {\bf 392}~(1802), 45--57.

\bibitem[Budanur {\em et~al.\/}(2014)Budanur, Cvitanovi{\'c}, Davidchack \&
  Siminos]{Budanur2014}
{\sc Budanur, N.~B., Cvitanovi{\'c}, P., Davidchack, R.~L. \& Siminos, E.} 2014
  Reduction of so(2) symmetry for spatially extended dynamical systems.

\bibitem[Chandler \& Kerswell(2013)]{Chandler_Kerswell2013_Kolm}
{\sc Chandler, G.~J. \& Kerswell, R.~R.} 2013 Invariant recurrent solutions
  embedded in a turbulent two-dimensional kolmogorov flow. {\em Journal of
  Fluid Mechanics\/} {\bf 722}, 554--595.

\bibitem[Cvitanovi{\'c}(2013)]{CvitanovicJFM2013_clockwork}
{\sc Cvitanovi{\'c}, Predrag} 2013 Recurrent flows: the clockwork behind
  turbulence. {\em Journal of Fluid Mechanics\/} {\bf 726}, 1--4.

\bibitem[Cvitanovi{\'c} {\em et~al.\/}(2013)Cvitanovi{\'c}, Artuso, Mainieri,
  Tanner \& Vattay]{ChaosBook}
{\sc Cvitanovi{\'c}, P., Artuso, R., Mainieri, R., Tanner, G. \& Vattay, G.}
  2013 Chaos: Classical and quantum.

\bibitem[Cvitanovic \& Eckhardt(1991)]{CvitanovicPOT_1991}
{\sc Cvitanovic, P. \& Eckhardt, B.} 1991 Periodic orbit expansions for
  classical smooth flows. {\em Journal of Physics A: Mathematical and
  General\/} {\bf 24}~(5), L237.

\bibitem[Del~{\'A}lamo \& Jimenez(2009)]{Alamo_JImenez2009}
{\sc Del~{\'A}lamo, J.~C. \& Jimenez, J.} 2009 Estimation of turbulent
  convection velocities and corrections to taylor's approximation. {\em Journal
  of Fluid Mechanics\/} {\bf 640}, 5--26.

\bibitem[Faisst \& Eckhardt(2003)]{FaisstEckhardt}
{\sc Faisst, H. \& Eckhardt, B.} 2003 Traveling waves in pipe flow. {\em Phys.
  Rev. Lett.\/} {\bf 91}~(22), 224502.

\bibitem[Fedele(2014{\natexlab{{\em a\/}}})]{FedeleEPL2014}
{\sc Fedele, F.} 2014{\natexlab{{\em a\/}}} Geometric phases of water waves.

\bibitem[Fedele(2014{\natexlab{{\em b\/}}})]{JFMFedele2014}
{\sc Fedele, Francesco} 2014{\natexlab{{\em b\/}}} On certain properties of the
  compact zakharov equation. {\em Journal of Fluid Mechanics\/} {\bf 748},
  692--711.

\bibitem[Froehlich \& Cvitanovi{\'c}(2012)]{Froehlich}
{\sc Froehlich, S. \& Cvitanovi{\'c}, P.} 2012 Reduction of continuous
  symmetries of chaotic flows by the method of slices. {\em Communications in
  Nonlinear Science and Numerical Simulation\/} {\bf 17}~(5), 2074 -- 2084.

\bibitem[Garrison \& Chiao(1988)]{Garrison_GeomPhases}
{\sc Garrison, J.~C. \& Chiao, R.~Y.} 1988 Geometrical phases from global gauge
  invariance of nonlinear classical field theories. {\em Phys. Rev. Lett.\/}
  {\bf 60}, 165--168.

\bibitem[Gibson {\em et~al.\/}(2008)Gibson, Halcrow \&
  Cvitanovi{\'c}]{GibsonetalJFMCouette2008}
{\sc Gibson, J.~F., Halcrow, J. \& Cvitanovi{\'c}, P.} 2008 Visualizing the
  geometry of state space in plane couette flow. {\em Journal of Fluid
  Mechanics\/} {\bf 611}, 107--130.

\bibitem[Gritsun(2011)]{Gritsun2011}
{\sc Gritsun, A.} 2011 Connection of periodic orbits and variability patterns
  of circulation for the barotropic model of atmospheric dynamics. {\em Doklady
  Earth Sciences\/} {\bf 438}~(1), 636--640.

\bibitem[Gritsun(2013)]{Gritsun2013}
{\sc Gritsun, A.} 2013 Statistical characteristics, circulation regimes and
  unstable periodic orbits of a barotropic atmospheric model. {\em
  Philosophical Transactions of the Royal Society A: Mathematical, Physical and
  Engineering Sciences\/} {\bf 371}~(1991).

\bibitem[Hannay(1985)]{Hannay}
{\sc Hannay, J~H} 1985 Angle variable holonomy in adiabatic excursion of an
  integrable hamiltonian. {\em Journal of Physics A: Mathematical and
  General\/} {\bf 18}~(2), 221.

\bibitem[Hopf(1931)]{Hopf1931}
{\sc Hopf, Heinz} 1931 {\"U}ber die abbildungen der dreidimensionalen
  sph{\"a}re auf die kugelfl{\"a}che. {\em Mathematische Annalen\/} {\bf
  104}~(1), 637--665.

\bibitem[Husem{\"o}ller(1994)]{Husemoller}
{\sc Husem{\"o}ller, Dale} 1994 {\em Fibre Bundles\/}, 3rd edn. {\em Graduate
  Texts in Mathematics, Book 20\/} 1. New York, Springer.

\bibitem[Kreilos {\em et~al.\/}(2014)Kreilos, Zammert \&
  Eckhardt]{KreilosJFM2014}
{\sc Kreilos, T., Zammert, S. \& Eckhardt, B.} 2014 Comoving frames and
  symmetry-related motions in parallel shear flows. {\em Journal of Fluid
  Mechanics\/} {\bf 751}, 685--697.

\bibitem[Krogstad {\em et~al.\/}(1998)Krogstad, Kaspersen \&
  Rimestad]{KrogstadPoF}
{\sc Krogstad, P.-{\AA}., Kaspersen, J.~H. \& Rimestad, S.} 1998 Convection
  velocities in a turbulent boundary layer. {\em Physics of Fluids\/} {\bf
  10}~(4), 949--957.

\bibitem[Pancharatnam(1956)]{Pancharatnam}
{\sc Pancharatnam, S.} 1956 Generalized theory of interference, and its
  applications. {\em Proceedings of the Indian Academy of Sciences - Section
  A\/} {\bf 44}~(5), 247--262.

\bibitem[Rowley {\em et~al.\/}(2003)Rowley, Kevrekidis, Marsden \&
  K.]{RowleyMarsden2003}
{\sc Rowley, C.~W., Kevrekidis, I.~G., Marsden, J.~E. \& K., Lust} 2003
  Reduction and reconstruction for self-similar dynamical systems. {\em
  Nonlinearity\/} {\bf 16}~(4), 1257.

\bibitem[Rowley \& Marsden(2000)]{RowleyMarsden2000}
{\sc Rowley, Clarence~W. \& Marsden, Jerrold~E.} 2000 Reconstruction equations
  and the karhunen--lo{\`e}ve expansion for systems with symmetry. {\em Physica
  D: Nonlinear Phenomena\/} {\bf 142}~(1--2), 1 -- 19.

\bibitem[Shapere \& Wilczek(1989)]{Shapere1}
{\sc Shapere, Alfred \& Wilczek, Frank} 1989 Geometry of self-propulsion at low
  reynolds number. {\em Journal of Fluid Mechanics\/} {\bf 198}, 557--585.

\bibitem[Siminos \& Cvitanovi{\'c}(2011)]{Siminos2011}
{\sc Siminos, E. \& Cvitanovi{\'c}, P.} 2011 Continuous symmetry reduction and
  return maps for high-dimensional flows. {\em Physica D: Nonlinear
  Phenomena\/} {\bf 240}~(2), 187 -- 198.

\bibitem[Simon(1983)]{Simon}
{\sc Simon, Barry} 1983 Holonomy, the quantum adiabatic theorem, and berry's
  phase. {\em Phys. Rev. Lett.\/} {\bf 51}~(24), 2167--2170.

\bibitem[Steenrod(1999)]{Steenrod}
{\sc Steenrod, Norman} 1999 {\em The Topology of Fibre Bundles\/}. Princeton,
  University Press.

\bibitem[Taylor(1938)]{Taylor1938}
{\sc Taylor, G.~I.} 1938 The spectrum of turbulence. {\em Proceedings of the
  Royal Society of London. Series A - Mathematical and Physical Sciences\/}
  {\bf 164}~(919), 476--490.

\bibitem[Tian \& Roberts(2003)]{TianRoberts2003}
{\sc Tian, X. \& Roberts, P.~J.W.} 2003 A 3d lif system for turbulent buoyant
  jet flows. {\em Experiments in Fluids\/} {\bf 35}~(6), 636--647.

\bibitem[Viswanath(2007)]{VISWANATH2007}
{\sc Viswanath, D.} 2007 Recurrent motions within plane couette turbulence.
  {\em Journal of Fluid Mechanics\/} {\bf 580}, 339--358.

\bibitem[Wedin \& Kerswell(2004)]{WedinKerswell2004}
{\sc Wedin, H. \& Kerswell, R.~R.} 2004 Exact coherent structures in pipe flow:
  travelling wave solutions. {\em Journal of Fluid Mechanics\/} {\bf 508},
  333--371.

\bibitem[Willis {\em et~al.\/}(2013)Willis, Cvitanovi{\'c} \&
  Avila]{WillisCvitanovic2013}
{\sc Willis, A.~P., Cvitanovi{\'c}, P. \& Avila, M.} 2013 Revealing the state
  space of turbulent pipe flow by symmetry reduction. {\em Journal of Fluid
  Mechanics\/} {\bf 721}, 514--540.

\end{thebibliography}

\end{document}